\documentclass[a4paper,11pt]{article}
\usepackage{wrapfig}
\usepackage{pos}
\def\MSbar{\overline{\rm MS\kern-0.5pt}\kern0.5pt}

\title{QCD mesonic screening masses up to high temperatures}

\author[a]{Mattia Dalla Brida}
\author[b,c]{Leonardo Giusti}
\author[d]{Tim Harris}
\author[b,c]{Davide Laudicina}
\author*[c]{Michele Pepe}

\affiliation[a]{Theoretical Physics Department, CERN,\\
  1211 Geneva 23, Switzerland}

\affiliation[b]{Dipartimento di Fisica, Universit\`a di Milano-Bicocca,\\
  Piazza della Scienza 3, I-20126 Milano, Italy}

\affiliation[c]{INFN, Sezione di Milano-Bicocca,\\
  Piazza della Scienza 3, I-20126 Milano, Italy}

\affiliation[d]{School of Physics and Astronomy, University of Edinburgh, \\
  Edinburgh EH9 3JZ, UK}

\emailAdd{mattia.dalla.brida@cern.ch}
\emailAdd{Leonardo.Giusti@mib.infn.it}
\emailAdd{tharris@ed.ac.uk}
\emailAdd{d.laudicina1@campus.unimib.it}
\emailAdd{pepe@mib.infn.it}

\abstract{
   We discuss a strategy to study non-perturbatively QCD up to very high temperatures by Monte Carlo simulations on the lattice. It allows
   not only the thermodynamic properties of the theory but also other interesting thermal features to be investigated. As a first concrete
   application, we compute the flavour non-singlet mesonic screening masses and we present the results of Monte Carlo simulations at 12
   temperatures covering the range from T $\sim$ 1 GeV up to $\sim$ 160 GeV in the theory with three massless quarks. On the one side, chiral symmetry
   restoration manifests itself in our results through the degeneracy of the vector and the axial vector channels and of the scalar and the
   pseudoscalar ones, and, on the other side, we observe a clear splitting between the vector and the pseudoscalar screening masses up to the
   highest investigated temperature. A comparison with the high-temperature effective theory shows that the known one-loop order in the
   perturbative expansion does not provide a satisfactory description of the non-perturbative data up to the highest temperature considered.

   \vspace{1cm}\hspace{10.5cm}
     CERN-TH-2022-198
}

\FullConference{%
  41st International Conference on High Energy physics - ICHEP2022\\
  6-13 July, 2022\\
  Bologna, Italy
}

\begin{document}
\maketitle

\section{Introduction}

When QCD is studied at finite temperature $T$, a new energy scale influences the dynamics. In the low temperature regime, we have
hadrons while when we consider very high temperatures we have a plasma of quarks and gluons that interact weakly.
The understanding of high-temperature QCD is based on the idea that, when $T$ is large, the dynamics is ruled by this energy scale and, due
to asymptotic freedom, the gauge coupling becomes small. In those conditions, QCD behaves in practice as a 3-dimensional system. The gluon sector is given by a
3-dimensional Yang-Mills theory with gauge coupling $g^2 T$ involving the spatial components of the gauge field; the temporal component
becomes an adjoint scalar field with a mass of the order of $g T$. The quarks pick up a mass proportional to $T$ and are very heavy
fields. Hence, taking this effective approach, we encounter three energy scales which, when the temperature is extremely high,
are well separated, $g^2 T/\pi \ll g T \ll \pi T$. In this setup the quarks are practically external sources and the main dynamics is related to the
3-dimensional Yang-Mills theory: thus, although the gauge coupling is weak, there are infrared non-perturbative effects that can be
relevant~\cite{Ginsparg:1980ef,Appelquist:1981vg,Braaten:1995jr}.\\
\indent
First-principles non-perturbative information can be obtained studying QCD on the lattice however this is numerically challenging in the regime
of high temperatures. On the one side, the state-of-the-art method to investigate QCD thermodynamics is based on the direct measurement of
the trace anomaly of the Energy-Momentum tensor from which one can then obtain the other thermodynamics quantities. This method requires to consider always at the
same time two different energy scales and the maximal temperature that has been reached is about 2 GeV. A second problem is related to
the lack of knowledge of the parameters to use for the numerical simulations. For example, mesonic screening masses have been measured only
up to about 1 GeV in the continuum limit~\cite{Cheng:2010fe,Brandt:2014uda,Bazavov:2019www,Brandt:2019ksy}. Taking into account the inverse
logarithmic behaviour of the gauge coupling with the temperature, the range of the explored temperatures is quite limited.  \\
\indent
The first problem has been solved by studying thermodynamics in a moving reference
frame~\cite{Giusti:2011kt,Giusti:2010bb,Giusti:2012yj}: this has the great advantage that one 
has to perform Monte Carlo simulations only at the physical temperature of interest. Using this approach, the
thermodynamic features of the $SU(3)$ Yang-Mills theory have been measured very accurately up to high
temperatures~\cite{Giusti:2014ila,Giusti:2016iqr} and work is in 
 progress in QCD. For what concerns the lines of constant physics -- namely, the parameter to consider in the Monte Carlo simulations
 --  renormalizing the theory with a hadronic scheme is not a convenient choice since one has to consider a very fine lattice with a large
 size. A better choice is based on a finite volume scheme~\cite{Luscher:1991wu,Jansen:1995ck} where, by performing successive steps in which
 the energy is doubled and the lattice spacing is halved, one can reach very high energies with a fine lattice spacing. Exploiting this
 strategy, one can investigate QCD at high temperatures and, as a first application, we have measured the mesonic screening masses~\cite{DallaBrida:2021ddx}.

\section{Mesonic screening masses}

Mesonic screening masses $m_{{\cal O}} $ characterize the long distance behaviour of mesonic two-point functions in the spatial direction
\begin{equation}\label{eq:2pt}
  C_{{\cal O}}(x_3) =\int d x_0 dx_1\, dx_2\, \langle {\cal O}^a (x) {\cal O}^a (0) \rangle 
  \underset{x_3\rightarrow \infty}{\sim} e^{-m_{{\cal O}} x_3}
\end{equation}
\noindent
where ${{\cal O}} ^a (x) = \bar{\psi}(x) \, \Gamma \, T_a \, \psi(x) $, is a bilinear operator
with $T_a$ being the traceless generators of the flavour group and $\Gamma=1\!\!1, \gamma_5, \gamma_\mu,\gamma_\mu\gamma_5$ correspond to the
scalar, pseudoscalar, vector and axial-vector screening masses respectively. They give information on the response of the system to the
insertion of a mesonic operator as well as on the restoration of chiral symmetry. Screening masses are quite simple to measure
numerically and we can also compare the non-perturbative results with the perturbative calculation that has
been performed in the framework of the effective theory at one-loop order. For 3 flavours, it is given by~\cite{Laine:2003bd}
\begin{equation}\label{eq:masspt}
m_{{\cal O}}^{PT} = 2 \pi T \left( 1 + 0.032739961 g^2 \right).
\end{equation}
Notice that, at leading order, the mass is given by twice the mass of a static quark at finite temperature, $\pi T$, and that, at one-loop
order, all the mesonic screening masses are degenerate.

\section{The numerical study}

We compute non-perturbatively the mesonic screening masses at 12 values of the temperature, $T_0, \ldots, T_{11}$, approximately equally
spaced in logarithmic scale from 160~GeV to 1~GeV. We regularize QCD on the lattice with 3 flavours of massless clover fermions; for the
gauge sector we consider the Wilson action for $T_0, \ldots, T_8$ and the tree-level improved Symanzik 
action for $T_9, T_{10}$ and $T_{11}$.

Monte Carlo simulations are carried out on lattices with spatial size $L/a=288$ and $L\,T>20$ so to make finite size effects negligible. At each
physical temperature, 3 or 4 different values of the lattice spacing $a$ have been considered in order to
extrapolate the results to the continuum limit; only for the temperature $T_0$, the continuum limit has been obtained based on 2 values of
the lattice spacing. We have run our numerical simulations considering shifted boundary conditions with shift $\xi= (1,0,0)$ since we
have observed~\cite{Giusti:2016iqr} that lattice artifacts are particularly small for that choice. 

\begin{figure}[th]
  \begin{center}
\includegraphics[width=0.9\textwidth]{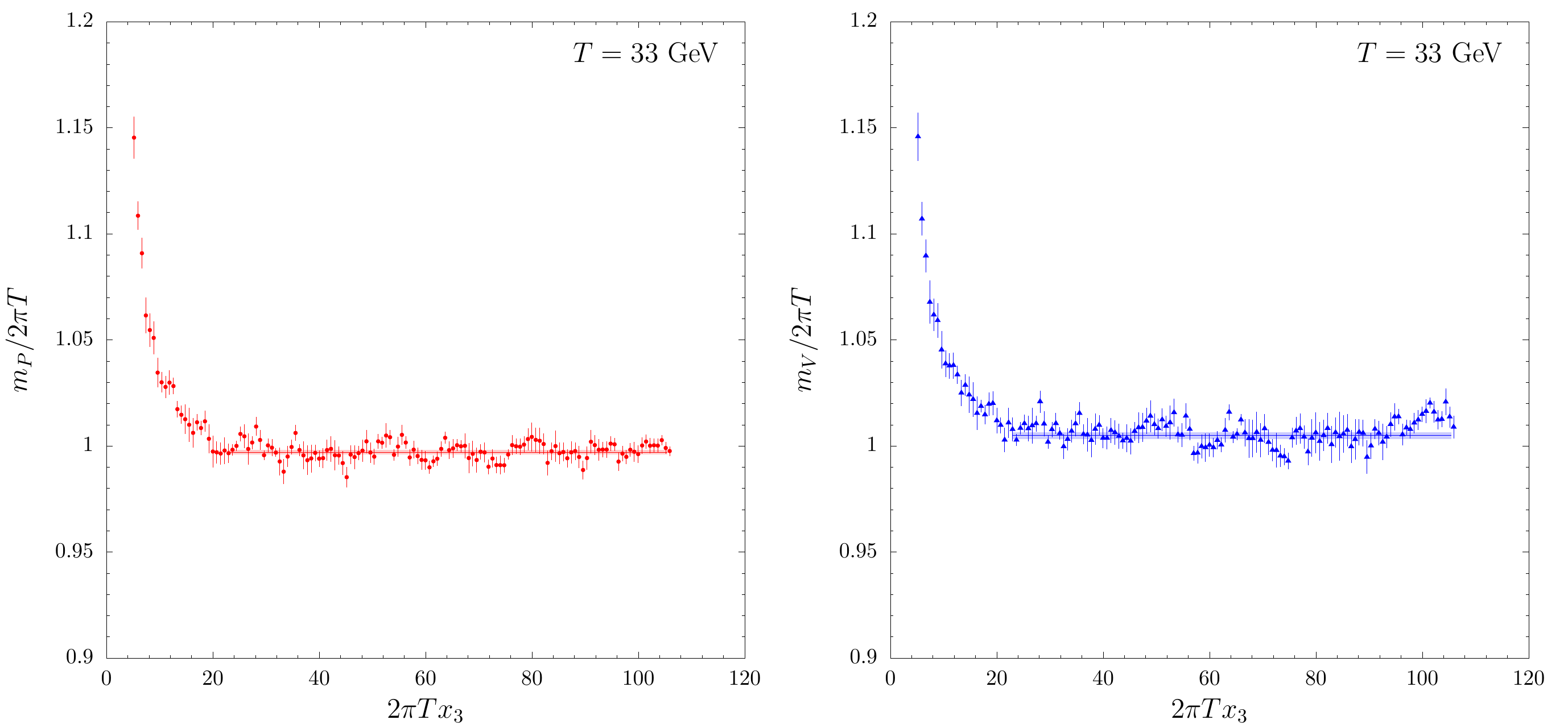}
\caption{Plot of the effective masses, normalized to $2\pi T$, for the pseudoscalar (left) and vector (right) correlators at
$T=33$ GeV for $L_0/a=6$. The two bands represent the estimate of the screening masses.\label{fig:plateau}}
\end{center}
\end{figure}

Effective masses are extracted from the expectation value of the two-point function $C_{{\cal O}}$ by
\begin{equation}
  m_{{\cal O}} (x_3)=\frac{1}{a} \, \mbox{arccosh} \left[ \frac{ C_{{\cal O}}(x_3+a) + C_{{\cal O}}(x_3-a)}{2 C_{{\cal O}}(x_3)} \right] 
\end{equation}
and the screening mass $m_{{\cal O}}$ is obtained as the value at which the effective mass flattens when $x_3$ becomes large. In
figure~\ref{fig:plateau} we show the effective mass plots for the pseudoscalar and the vector screening masses at the temperature $T_3$=33
GeV obtained for a lattice with temporal extent $L_0/a=6$. In both cases, we notice long plateaux that allow us to measure the screening
masses with high accuracy and that are shown as bands in the plots.

At all the temperatures we have investigated, we observe a degeneracy between the scalar
and the pseudoscalar screening masses as well as between the vector and the axial-vector ones: this is evidence for the restoration of chiral
symmetry at high temperatures. More precisely, the $U_A(1)$ symmetry is anomalous also at finite temperature but the very strong suppression
of the topological susceptibility taking place at high temperatures, makes the effects of the anomaly negligible w.r.t. the statistical
uncertainty. For this reason we focus our discussion on the pseudoscalar, $m_P$, and on the vector, $m_V$, screening masses.

\begin{figure}[ht!]
\begin{center}
\includegraphics[width=0.45\textwidth]{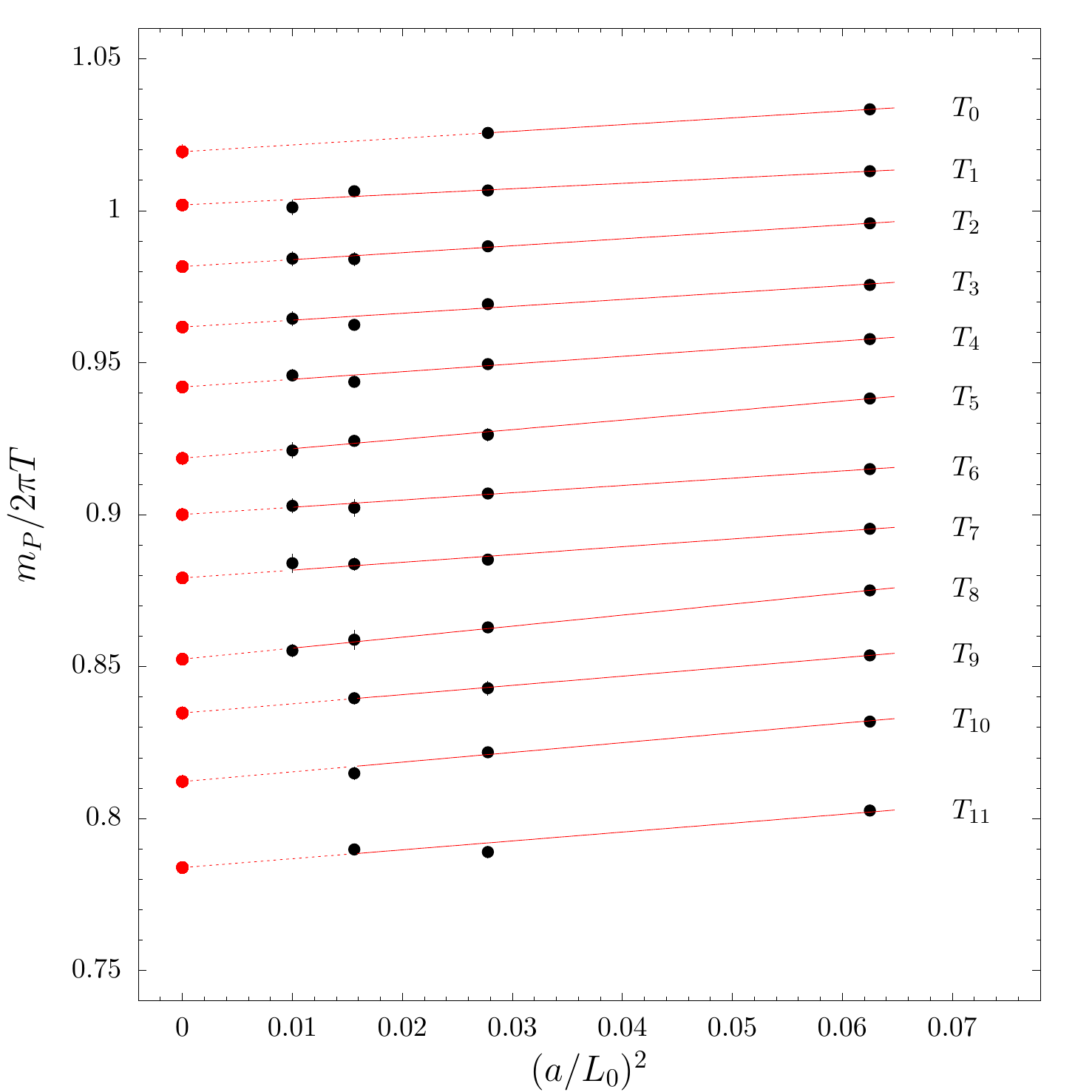}
\includegraphics[width=0.45\textwidth]{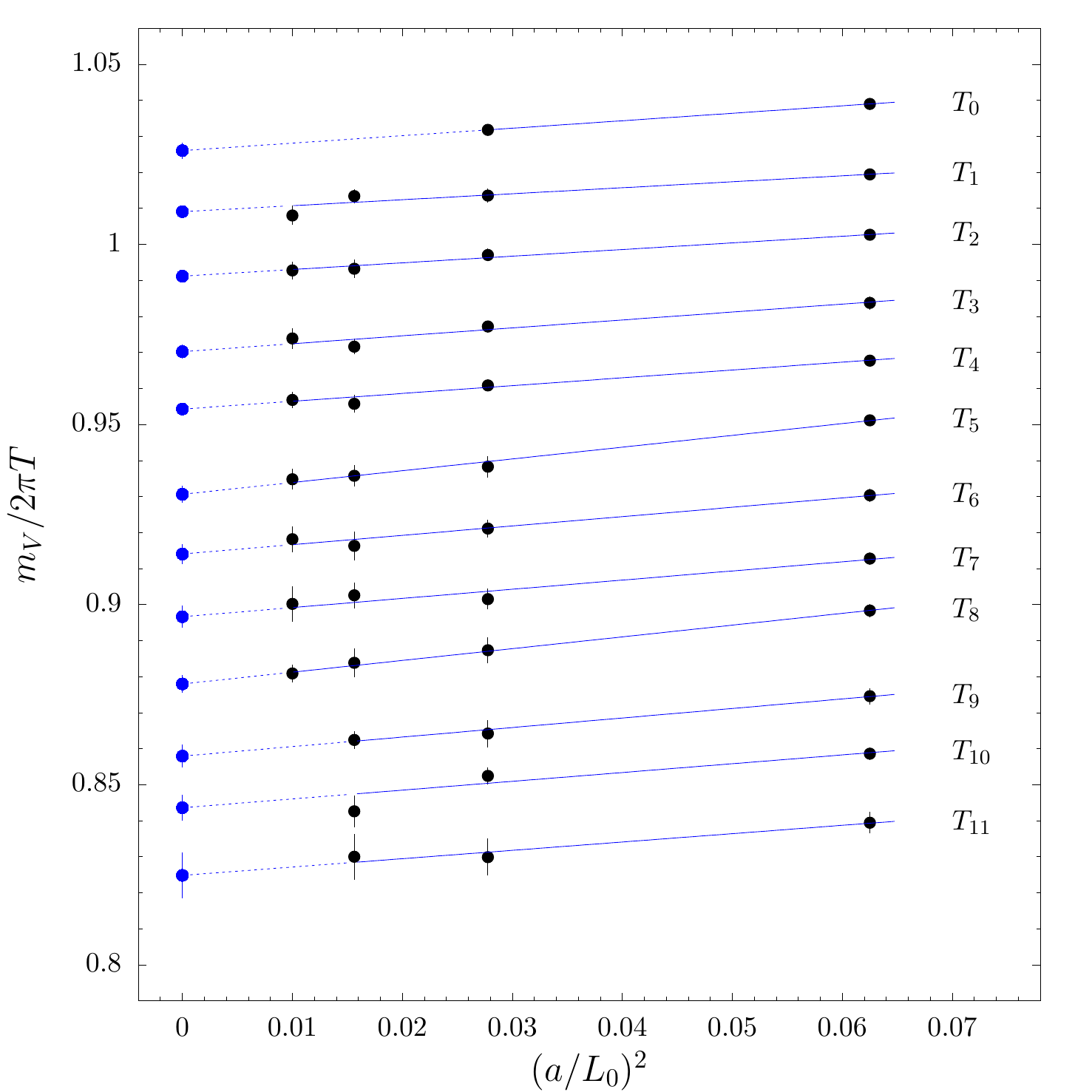}
  \caption{Continuum limit extrapolations for the tree-level improved pseudoscalar (left panel) and vector (right panel) screening masses. Each temperature
  is analyzed independently from the others. Data corresponding to $T_i$ ($i=0,\dots,11$) are shifted downward by $0.02\times i$ for better readability.
\label{fig:PV_extrCL}}
\end{center}
\end{figure}

In order to ameliorate the approach to the continuum limit, we have introduced the following tree-level improved definiton of the screening
mass 
\begin{equation}
m_{{\cal O}} \longrightarrow m_{{\cal O}} - \Big[m^{\rm free}_{{\cal O}} - 2\pi T\Big]\; , 
\end{equation}
where $m^{\rm free}_{{\cal O}}$ is the screening mass computed in the free lattice theory. In figure~\ref{fig:PV_extrCL}
we show the continuum limit extrapolations at the 12 temperatures we have considered. We notice that the $O(a)$-improved fermion action, the
improved definition of the screening mass and the use of shifted boundary conditions with shift vector $\xi= (1,0,0)$ make the
discretization effects very mild and we attain a final accuracy in the continuum limit of a few permille.

\begin{wrapfigure}{r}{7.5cm}
  \centering
\includegraphics[width=0.49\textwidth]{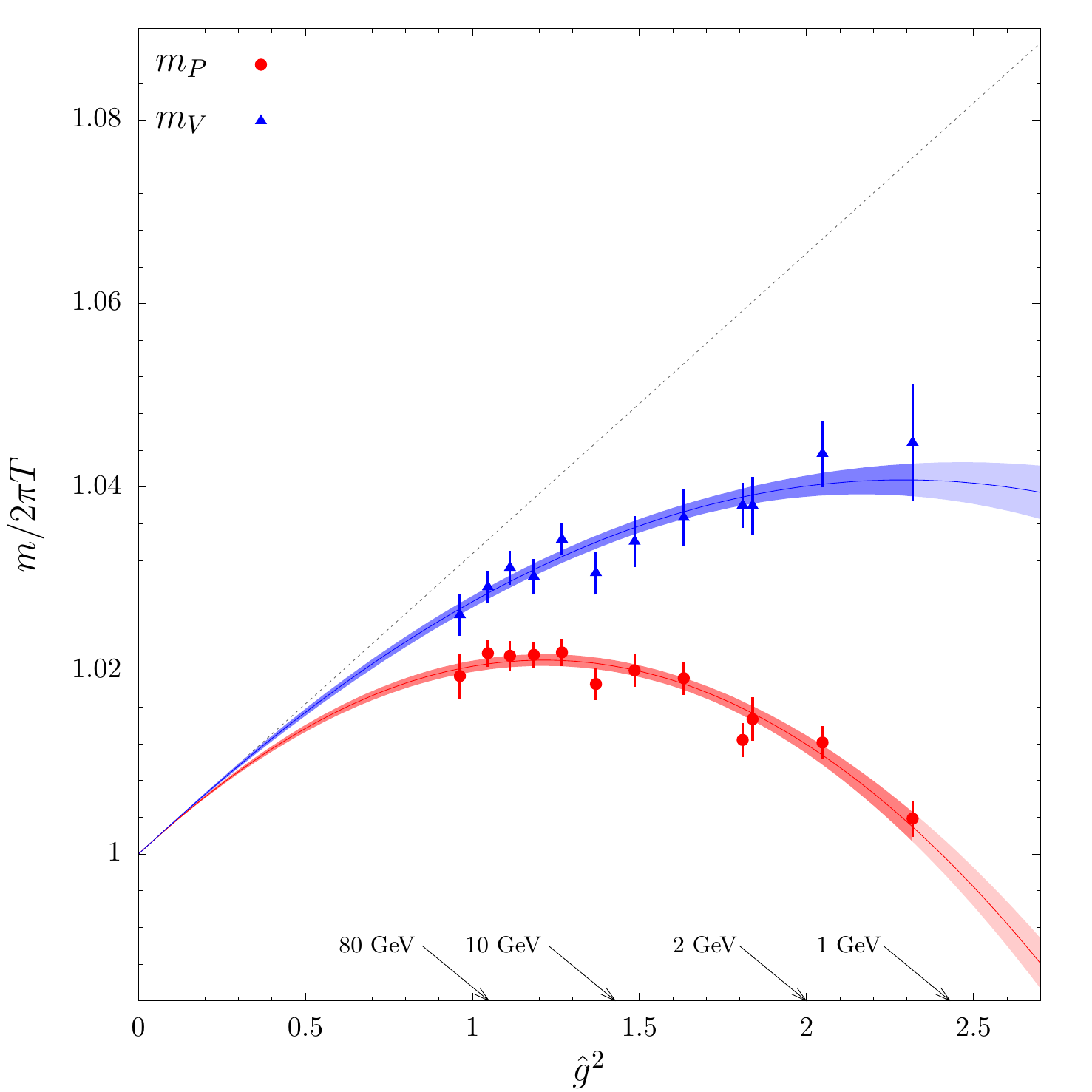}      
\caption{Pseudoscalar (red) and vector (blue) screening masses versus $\hat g^2$. The bands
         represent the best fits up to order $\hat g^4$, while the
         dashed line is eq.~(\ref{eq:masspt}).\label{fig:massCL}}
\end{wrapfigure}

In figure~\ref{fig:massCL} we present the temperature dependence of the pseudoscalar and of the vector screening masses: the data are those resulting from the continuum limit extrapolations of figure~\ref{fig:PV_extrCL}. 
The straight line represents the one-loop perturbative result given by eq.~(\ref{eq:masspt}). The dependence on the temperature
is expressed in terms of the function $\hat g^2 (T)$ defined as  
\begin{equation}\label{eq:gmu}
  \frac{1}{\hat g^2(T)} \equiv \frac{9}{8\pi^2} \ln  \frac{2\pi T}{\Lambda_{\MSbar}} 
  + \frac{4}{9 \pi^2} \ln \left( 2 \ln  \frac{2 \pi T}{\Lambda_{\MSbar}}  \right)\; , 
\end{equation}
where $\Lambda_{\MSbar} = 341$~MeV is taken from Ref.~\cite{Bruno:2017gxd}. Although it corresponds to the two-loop $\MSbar$ definition of the
strong coupling constant at the scale $\mu=2\pi T$, for our purposes it is just a function
of the temperature $T$, suggested by the effective theory analysis, that we use to analyze our results: the crucial point is the leading
logarithmic dependence on $T$. The blue and the red bands are fits of the numerical data up to order $\hat g^4 (T)$. We refer the reader
to~\cite{DallaBrida:2021ddx} for a detailed discussion of the data analysis.

\section{Conclusion}
In these proceedings we have presented a strategy to investigate QCD up to very high temperatures. On the one side, we consider shifted
boundary conditions that represent an effective framework to investigate thermodynamics (and for $\xi= (1,0,0)$ give, in general, very
small lattice artifacts) and, on the other side, we match $T$ with the energy scale of the running coupling defined with the step-scaling
technique in order to determine the parameters for the Monte Carlo simulations. We show results for mesonic screening masses at temperatures
ranging from 1 GeV to 160 GeV and we compare our non-perturbative results with the one-loop perturbative computation: contrary to the
expectation coming from the latter, we observe a splitting between pseudoscalar and vector screening masses for all the investigated
temperatures. We find also evidence for a degeneracy between the scalar and the pseudoscalar masses as well as between the vector and the 
axial-vector ones.\\
{\it \underline {Acknowledgements}}
We acknowledge PRACE for awarding us access to the HPC system MareNostrum4 at the Barcelona Supercomputing Center
(Proposals n. 2018194651 and 2021240051). We also thank CINECA for providing us with computer-time on Marconi (CINECA- INFN, CINECA-Bicocca
agreements, ISCRA B projects HP10BF2OQT and HP10B1TWRR).

\bibliographystyle{JHEP}
\bibliography{bibfile.bib}

\providecommand{\href}[2]{#2}\begingroup\raggedright\begin{thebibliography}{10}

\bibitem{Ginsparg:1980ef}
P.~H. Ginsparg, {\it {First Order and Second Order Phase Transitions in Gauge
  Theories at Finite Temperature}},  {\em Nucl. Phys. B} {\bf 170} (1980)
  388--408.

\bibitem{Appelquist:1981vg}
T.~Appelquist and R.~D. Pisarski, {\it {High-Temperature Yang-Mills Theories
  and Three-Dimensional Quantum Chromodynamics}},  {\em Phys. Rev. D} {\bf 23}
  (1981) 2305.

\bibitem{Braaten:1995jr}
E.~Braaten and A.~Nieto, {\it {Free energy of QCD at high temperature}},  {\em
  Phys. Rev. D} {\bf 53} (1996) 3421--3437,
  [\href{http://arxiv.org/abs/hep-ph/9510408}{{\tt hep-ph/9510408}}].

\bibitem{Cheng:2010fe}
M.~Cheng et~al., {\it {Meson screening masses from lattice QCD with two light
  and the strange quark}},  {\em Eur. Phys. J. C} {\bf 71} (2011) 1564,
  [\href{http://arxiv.org/abs/1010.1216}{{\tt arXiv:1010.1216}}].

\bibitem{Brandt:2014uda}
B.~B. Brandt, A.~Francis, M.~Laine, and H.~B. Meyer, {\it {A relation between
  screening masses and real-time rates}},  {\em JHEP} {\bf 05} (2014) 117,
  [\href{http://arxiv.org/abs/1404.2404}{{\tt arXiv:1404.2404}}].

\bibitem{Bazavov:2019www}
A.~Bazavov et~al., {\it {Meson screening masses in (2+1)-flavor QCD}},  {\em
  Phys. Rev. D} {\bf 100} (2019), no.~9 094510,
  [\href{http://arxiv.org/abs/1908.09552}{{\tt arXiv:1908.09552}}].

\bibitem{Brandt:2019ksy}
B.~B. Brandt, O.~Philipsen, M.~C\`e, A.~Francis, T.~Harris, H.~B. Meyer, and
  H.~Wittig, {\it {Testing the strength of the $\text{U}_A(1)$ anomaly at the
  chiral phase transition in two-flavour QCD}},  {\em PoS} {\bf CD2018} (2019)
  055, [\href{http://arxiv.org/abs/1904.02384}{{\tt arXiv:1904.02384}}].

\bibitem{Giusti:2011kt}
L.~Giusti and H.~B. Meyer, {\it {Thermodynamic potentials from shifted boundary
  conditions: the scalar-field theory case}},  {\em JHEP} {\bf 11} (2011) 087,
  [\href{http://arxiv.org/abs/1110.3136}{{\tt arXiv:1110.3136}}].

\bibitem{Giusti:2010bb}
L.~Giusti and H.~B. Meyer, {\it {Thermal momentum distribution from path
  integrals with shifted boundary conditions}},  {\em Phys. Rev. Lett.} {\bf
  106} (2011) 131601, [\href{http://arxiv.org/abs/1011.2727}{{\tt
  arXiv:1011.2727}}].

\bibitem{Giusti:2012yj}
L.~Giusti and H.~B. Meyer, {\it {Implications of Poincare symmetry for thermal
  field theories in finite-volume}},  {\em JHEP} {\bf 01} (2013) 140,
  [\href{http://arxiv.org/abs/1211.6669}{{\tt arXiv:1211.6669}}].

\bibitem{Giusti:2014ila}
L.~Giusti and M.~Pepe, {\it {Equation of state of a relativistic theory from a
  moving frame}},  {\em Phys. Rev. Lett.} {\bf 113} (2014) 031601,
  [\href{http://arxiv.org/abs/1403.0360}{{\tt arXiv:1403.0360}}].

\bibitem{Giusti:2016iqr}
L.~Giusti and M.~Pepe, {\it {Equation of state of the SU(3) Yang–Mills
  theory: A precise determination from a moving frame}},  {\em Phys. Lett.}
  {\bf B769} (2017) 385--390, [\href{http://arxiv.org/abs/1612.00265}{{\tt
  arXiv:1612.00265}}].

\bibitem{Luscher:1991wu}
M.~L{\"u}scher, P.~Weisz, and U.~Wolff, {\it {A numerical method to compute the
  running coupling in asymptotically free theories}},  {\em Nucl. Phys.} {\bf
  B359} (1991) 221--243.

\bibitem{Jansen:1995ck}
K.~Jansen et~al., {\it {Non-perturbative renormalization of lattice QCD at all
  scales}},  {\em Phys. Lett.} {\bf B372} (1996) 275--282,
  [\href{http://arxiv.org/abs/hep-lat/9512009}{{\tt hep-lat/9512009}}].

\bibitem{DallaBrida:2021ddx}
M.~Dalla~Brida, L.~Giusti, T.~Harris, D.~Laudicina, and M.~Pepe, {\it
  {Non-perturbative thermal QCD at all temperatures: the case of mesonic
  screening masses}},  {\em JHEP} {\bf 04} (2022) 034,
  [\href{http://arxiv.org/abs/2112.05427}{{\tt arXiv:2112.05427}}].

\bibitem{Laine:2003bd}
M.~Laine and M.~Vepsalainen, {\it {Mesonic correlation lengths in high
  temperature QCD}},  {\em JHEP} {\bf 02} (2004) 004,
  [\href{http://arxiv.org/abs/hep-ph/0311268}{{\tt hep-ph/0311268}}].

\bibitem{Bruno:2017gxd}
{\bf ALPHA} Collaboration, M.~Bruno, M.~Dalla~Brida, P.~Fritzsch, T.~Korzec,
  A.~Ramos, S.~Schaefer, H.~Simma, S.~Sint, and R.~Sommer, {\it {QCD Coupling
  from a Nonperturbative Determination of the Three-Flavor $\Lambda$
  Parameter}},  {\em Phys. Rev. Lett.} {\bf 119} (2017), no.~10 102001,
  [\href{http://arxiv.org/abs/1706.03821}{{\tt arXiv:1706.03821}}].

\end{thebibliography}\endgroup

\end{document}